\documentclass[reprint,eqsecnum,floats,aps,amsmath,amssymb,nofootinbib,prd,twocolumn, showpacs]{revtex4-1}
\usepackage{amsmath,amsthm,latexsym,amssymb,amsfonts}

\begin{document}
	
\title{Quantum corrections to the Mukhanov-Sasaki equations}
\author{Laura Castell\'o Gomar}
\email{laura.castello@iem.cfmac.csic.es}
\author{Guillermo A. Mena Marug\'an}
\email{mena@iem.cfmac.csic.es}
\affiliation{Instituto de Estructura de la Materia, IEM-CSIC, Serrano 121, 28006 Madrid, Spain}
\author{Mercedes Mart\'in-Benito}
\email{m.martin@hef.ru.nl}
\affiliation{Radboud University Nijmegen, Institute for Mathematics, Astrophysics and Particle Physics, \\Heyendaalseweg 135, NL-6525 AJ Nijmegen, The Netherlands}

\begin{abstract}
Recently, a lot of attention has been paid to the modifications of the power spectrum of primordial fluctuations caused by quantum cosmology effects. The origin of these modifications are corrections to the Mukhanov-Sasaki equations that govern the propagation of the primeval cosmological perturbations. The specific form of these corrections depends on a series of details of the quantization approach and of the prescription followed to implement it. Generally, nonetheless, the complexity of the theoretical quantum formulation is simplified in practice appealing to a semiclassical or effective  approximation, in order to perform concrete numerical computations. In this work, we introduce technical tools and design a procedure to deal with these quantum corrections beyond the most direct approximations employed so far in the literature. In particular, by introducing an interaction picture, we extract the quantum dynamics of the homogeneous geometry in absence of scalar field potential and inhomogeneities, dynamics that has been intensively studied and that can be integrated. The rest of our analysis focuses on the interaction evolution, putting forward methods to cope with it. The ultimate aim is to develop treatments that increase our ability to discriminate between the predictions of different quantization proposals for cosmological perturbations. 
\end{abstract}


\maketitle

\section{Introduction}\label{sec:Intro}

Observational cosmology is living a golden age, impulsed by the impressive technical developments achieved in the last decades. These developments have led to a new era that is known as ``precision cosmology'' \cite{precisioncosmology}, in which a reasonable number of cosmological parameters have been measured with a precision of a few percent for the first time in history \cite{Planckparameters}. This has allowed us to compare predictions of theoretical models for cosmology with observations, and falsify some of them. One of the most relevant observations is the measurement of the Cosmic Microwave Background (CMB) \cite{CMB}. This background encodes information about the first stages of the Universe, previous to the decoupling of the radiation and matter contents. The CMB appears as an almost uniform radiation background with a spectrum that corresponds to a nearly perfect black body. The small variations superimposed as anisotropies on this radiation (of a relative order of $10^{-5}$) reveal the fluctuations of the Primeval Universe, and their steadily improved analysis is opening new windows for the understanding of the early cosmological stages. In addition, the polarization of the CMB is expected to provide most valuable information, e.g. about the generation of primordial gravitational waves in our Universe.

The latest Planck 2015 analysis of the power spectrum of the CMB \cite{Planckparameters,Planck} found an excellent agreement with the theoretical predictions of Standard Cosmology \cite{cosmology} over a broad range of scales. Even so, there are indications of some possible tensions between those predictions and the observations \cite{Planckparameters,Planckanomalies}, especially at large angular scales. For instance, there seems to be a lack of power for multipoles with low number $l$, corresponding to those scales, as well as unexpected features around $l=30$ \cite{Planckanomalies}. Such anomalies in the temperature anisotropies have raised an especial interest in the community of cosmologists, since they may have originated from fundamental physical processes occurring in the first epochs of the Universe. However, it is very difficult to find a conclusive explanation to these anomalies for now, owing to the uncertainty in the measurements at large angular scales, arising among other things from the problem of cosmic variance (since the number of available independent modes is too small, because its number is roughly speaking almost proportional to the inverse of the angular scale \cite{cosmology}). Nonetheless, a very appealing possibility is that such anomalies and deviations might be the result of genuine quantum gravity effects (beyond the quantum fluctuations of fields in a classical cosmological spacetime) that would have affected the primordial cosmological perturbations and would have been imprinted on the primary CMB anisotropies. These ideas are transforming the field of quantum cosmology, boosting the extraction of predictions from quantum models of the Universe with (more than abstract) hopes of confronting them with the available observations and those expected in a near future.

Many of the studies about the consequences of quantum cosmology in observational astrophysics and, in particular, in the CMB have been carried out within the framework of Loop Quantum Cosmology (LQC) \cite{hybrid,hybr-inf3,javidani,dress,ivans,ivans2,deformedLQC,BojGian}. However, the discussion is not limited to that formalism by any means \cite{hybr-inf4}, and other formalisms like quantum geometrodynamics \cite{kiefermanuel,kieferdavidmanuel} have also been explored. Even within LQC, different alternate formulations have been suggested and investigated. This diversity in the lines of attack must be viewed as a reflection of the increasing interest paid to the study of realistic observational consequences in cosmology of the quantum nature of the spacetime. 

Let us focus our comments on LQC. LQC is the application of the methods of Loop Quantum Gravity (LQG) \cite{LQG} to cosmology. It has been successfully applied to the simple case of homogeneous and isotropic spacetimes: The so called Friedmann-Lema\^{\i}tre-Robertson-Walker (FLRW) cosmologies \cite{bounce,improved,LQCFLRW,LQC,AS,sigma}. One of the most spectacular predictions is the replacement of the big bang singularity by a quantum bounce \cite{bounce,improved}. The first inhomogeneous cosmological model studied within LQC was a family of spacetimes with two spatial isometries, known as the Gowdy model \cite{Gowdy}. The study was restricted to the case of the spatial topology of the three-torus $T^3$  and to gravitational waves with linear polarization. A hybrid quantization strategy was adopted, using loop methods for the zero-modes of the geometry, while Fock techniques were employed for the fields describing the inhomogeneities \cite{LQCGowdy}. The philosophy behind this strategy rests on the assumption that the most relevant aspects of the quantum nature of the geometry are encoded in zero-modes, whereas other modes can be described essentially along the lines of quantum field theory (QFT) in curved spacetimes. Therefore, it can be understood as a quantum dynamical regime in between that of full quantum gravity and a more conventional QFT in a fixed curved background. 

More recently, this hybrid quantization approach has been applied to the case of the Gowdy model with a scalar field as matter content \cite{Gowdymatter} and, more importantly for our discussion here, to cosmological perturbations in inflationary scenarios \cite{hybrid,hybr-inf3,hybr-inf4}. A careful analysis has been performed of the perturbations of an FLRW spacetime with a scalar field (which represents the so-called {\it inflaton}), minimally coupled and subject to a potential (a quadratic one in most of the calculations, for the sake of simplicity). This analysis considers perturbations in both the spacetime metric and the scalar field. According to the current cosmological model favored by observations \cite{Planckparameters}, the discussion has been particularized to the case of flat spatial topology, assuming for simplicity that is compact (namely, that of $T^3$) and bearing in mind that this assumption is innocuous as long as the compactification scale is large compared to the radius of the observable Universe \cite{hybrid}. 

Following this hybrid approach and using a kind of Born-Oppenheimer ansatz for physical states, it is possible to deduce the quantum equation that dictates the dynamics of the cosmological perturbations \cite{hybr-inf3,hybr-inf4}.  As long as one can neglect quantum transitions of the FLRW geometry, and the (quadratic) dependence on the rest of physical degrees of freedom admits a direct (classical) effective counterpart, one obtains effective equations for the inhomogeneities that, in a conformal time, are nothing but the Mukhanov-Sasaki (MS) equations \cite{MS} with corrections that take into account quantum gravitational effects \cite{hybr-inf3,hybr-inf4}. This modified MS equations provide the master equations to extract predictions about the power spectrum of the primordial perturbations. 

Similar modified MS equations have also been derived within LQC following an alternate proposal called ``dress metric'' approach \cite{dress}. In this approach, however, one renounces to treat the full system of the FLRW geometry plus the perturbations as a symplectic manifold,\footnote{This belief in the loss of a global symplectic structure seems to be postulated as well in Ref. \cite{kieferdavidmanuel}.}  even if both General Relativity with a scalar field and its truncation at quadratic perturbative order in the action are known to possess a phase space that indeed admits a symplectic structure \cite{hybr-inf4,BojGian,H-H}. In the dress metric approach, and exactly as it happens in the hybrid quantization with an appropriate Born-Oppenheimer approximation, the functions of the FLRW geometry that appear in the MS equations are replaced with expectation values on the physical state that describes that geometry. In this respect, the difference between the two approaches consists in details about which are those expectation values and about quantization prescriptions affecting the operators in those values. Let us also mention that the possible consequences of LQC in the CMB have been studied as well by adopting another viewpoint, namely, by postulating the deformations of the spacetime diffeomorphisms algebra that one might expect that arise in LQG, demanding then the closure of the modified algebra for consistency \cite{BojGian}. In this latter approach, the modifications are already deduced in the form of corrected effective equations for the perturbations.

In addition to all these studies, in a recent work we provided a full covariant description of the entire system corresponding to perturbed FLRW spacetimes with a scalar field (at our considered perturbative level) \cite{hybr-inf4}. In particular, the perturbative physical degrees of freedom were described by MS gauge invariants, and we found a complete canonical set by determining new background variables that are corrected with perturbative contributions in order to preserve the symplectic structure of the total system. In that work, and inspired by the analyses carried out in the hybrid approach to LQC, the derivation of the modified MS equations was generalized without specifying the quantization adopted for the zero-modes of the FLRW geometry \cite{hybr-inf4}. There is certain parallelism between this derivation and the discussion that has been presented even more recently for geometrodynamics in Ref. \cite{kieferdavidmanuel}.

Detailed investigations of the effects of these modifications on the MS equations have been performed only after introducing certain semiclassical or effective approximations for the description of the FLRW geometry. In the dress metric approach to LQC, the expectation values on the FLRW state have been replaced with the corresponding {\it classical} values evolved according to the effective equations of homogeneous and isotropic LQC \cite{effective,AS}. No trace of backreaction has been considered, but only cases in which this backreaction can be totally neglected have been contemplated. In this way, sectors of solutions with predictions compatible with most of the features of the CMB have been identified, and modifications to the scalar power spectrum have been unveiled in the region of large angular scales \cite{ivans,ivans2}. The most interesting of these modifications involves a loss of Gaussianity owing to the evolution of the vacuum state of the perturbations prior to the slow-roll phase \cite{ivans2}. This loss allows for correlations between observable modes and those beyond the Hubble horizon. Such correlations have been investigated by considering the three-point function of the perturbations, which should provide the most important correction. Nonetheless, in situations in which this correction is of the order of the Gaussian contribution, a more detailed quantum calculation would be desirable to determine not only the qualitative, but also the quantitative effect. Besides, the calculations have been performed assuming that the correlations are only important in the slow-roll regime \cite{ivans2}. In the hybrid approach to LQC, a careful numerical analysis of the corrections to the CMB have been carried out only very recently, in Ref. \cite{javidani}. Also in this case the calculation has assumed effective equations for the evolution of the FLRW geometry and negligible backreaction. In fact, probably the main difference with respect to previous studies in LQC is a new proposal for the choice of vacuum of the fluctuations, which strictly speaking is not a genuine quantum geometry effect. On the other hand, in Ref. \cite{kieferdavidmanuel}, the modifications to the CMB power spectrum in geometrodynamics have been discussed using a semiclassical (WKB) definition of time, as well as semiclassical values for functions of the FLRW geometry that appear in the corrected MS equations. A new proposal for the vacuum of the perturbations is also introduced. Most importantly, a de Sitter approximation is employed, ignoring terms arising from variations of the scalar field (which significantly complicate the computations). 

The aim of the present work is to provide formulas and methods for the computation of the modifications to the MS equations that keep the quantum corrections up to the maximum practical extent. In the context that we have discussed, this goal is important for several reasons. First, as we have explained, the predictions of various formalisms and prescriptions in quantum  cosmology (and in particular in LQC) can be satisfactorily discriminated only if quantum corrections beyond the semiclassical or effective approximations are maintained. Therefore, any hope of falsification by means of CMB observations would depend on this analysis. Second, the exploration of sectors of states in which the corrections to the CMB power spectrum are {\emph{important}} for certain ranges of scales makes it advisable that we improve our approximations if we want to reach better quantitative estimations. Third, the inclusion of more corrections coming from quantum geometry effects can reveal new phenomena in the CMB spectrum. In a certain sense, one might expect that this is also the case for the search of a new prescription for the vacuum of the perturbations, which should be formulated in a description that incorporates all those phenomena. And finally, clearly, this kind of analysis provides a manner to check whether the additional quantum corrections are indeed negligible in many situations where this has been assumed.

The main obstacle to compute the expectation values over the FLRW geometry that determine the modified MS equations is found in the quantum evolution of the FLRW states. This evolution is dictated by a quantum Hamiltonian which is not integrable in the presence of a non-constant potential for the scalar field \cite{hybr-inf3,hybr-inf4}. Moreover, there exist complications even for the numerical integration of this quantum evolution \cite{HKT}. The strategy that we propose in this paper to deal with the problem is the following. a) From the generator of the FRLW dynamics, extract its free geometric part, corresponding to a vanishing potential, and use it to pass to an interaction image. b) Integrate explicitly the evolution generated by this free geometric operator. In LQG, for instance, this is possible even analytically by adopting a prescription called ``solvable LQC'' (sLQC) \cite{sLQC}. We will focus our discussion on this prescription. c) From the generator of the interaction dynamics in FLRW, extract the dominant contributions of the scalar field potential, and pass to a new interaction image (in doing this, regard the potential as a kind of perturbation; for instance, for a quadratic potential, one can use the mass of the scalar field as a small parameter). d) Expand the evolution operator corresponding to the dominant contributions of the potential, keeping only those terms considered to be relevant. And e) Integrate (semi-)classically the effective evolution generated by the remaining interaction terms. Notice that only the two last points (d and e) imply approximations: In the former of these points, a quantum truncation, and in the latter, a semiclassical (or effective) approximation. However, the free geometric part of the FLRW dynamics, as well as the most relevant contributions of the potential, are treated exactly within quantum mechanics. 
 
The rest of the paper is organized as follows. In Sec. \ref{sec:sLQC} we briefly review the quantization of the FLRW spacetimes with a homogeneous scalar field in LQC and particularize the discussion to sLQC, determining the evolution in the absence of field potential. In Sec. \ref{sec:MS} we summarize the results of Ref. \cite{hybr-inf4}, present the modified MS equations, and particularize the analysis to the case of sLQC. We then start with our reformulation of the quantum corrections to the MS equations in Sec. \ref{sec:Interaction}, expressing the expectation values that appear in those equations in an interaction image. Later on, in Sec. \ref{sec:potential}, we extract the dominant contributions to the interaction and to its corresponding evolution operator, specializing the study to the case of a mass term. We complete our treatment in Sec. \ref{sec:Remaining}, determining the remaining dynamics and discussing its semiclassical or effective counterpart. We discuss our results and conclude in Sec. \ref{sec:Conclu}. Some particularly long formulas are included in an appendix.

\section{Homogeneous sector and sLQC}
\label{sec:sLQC}

In this section we will summarize the quantization of the zero-mode sector of the perturbed FLRW spacetimes with a scalar field, which can be regarded as the phase space of a homogeneous and isotropic model, obtained by unplugging the inhomogeneities of the system \cite{hybr-inf4}. We set the reduced Planck constant $\hbar$ and the speed of light equal to $1$, and the period of the orthogonal coordinates of the spatial sections $T^3$ equal to $2\pi$. Here, we will quantize this zero-mode sector following the prescriptions of sLQC. 

In LQG, the gravitational degrees of freedom are described by a real $su(2)$-connection, called the Ashtekar-Barbero connection, and by a densitized triad \cite{LQG}. The fundamental variables in LQG are {\it smeared} functions of these phase space coordinates, namely, holonomies of the connection and fluxes of the densitized triad. In models that are homogeneous and isotropic, the physical freedom in the densitized triad is just a global factor, usually denoted as $p$, which varies with time. The sign of $p$ depends on the orientation of the triad, and its absolute value is the squared scale factor of the FLRW spacetime up to a constant numerical factor \cite{LQC,AS,sigma}. The physical freedom in the connection, on the other hand, is also encoded by one single dynamical variable, called $c$. For flat topology, the connection is proportional to $c$. This variable is canonical to $p$ inasmuch as their Poisson bracket is $\{c,p\}=8\pi G \gamma/3$, where $G$ is the Newton constant and $\gamma$ is a constant known as the Immirzi parameter \cite{AS,Immirzi}. The basic holonomies of the connection are taken along straight lines with a length such that the square formed by them has a physical area equal to the non-vanishing minimum $\Delta$ allowed by LQG  \cite{area}. This prescription to choose the holonomies is called ``improved dynamics'' \cite{improved}. On the other hand, fluxes are just proportional to $p$. 

To simplify the  calculations, it is common to change variables from $c$ and $p$ to a new canonical set such that the chosen holonomies simply produce a constant shift of a unit in the new geometric variable that replaces $p$. The new set is
\begin{equation}\label{vp}
v={\rm sign}{(p)} \frac{|p|^{3/2}}{2\pi G \gamma  \sqrt{\Delta}},\quad b=\sqrt{\frac{\Delta}{|p|}}c,
\end{equation}
with $\{b, v\}=2$. The physical volume of the $T^3$ section of the FLRW spacetime is $V=2\pi G \gamma  \sqrt{\Delta}|v|$, proportional to the absolute value of $v$. 

The matter content of the sector is the zero-mode of the scalar field of the model. This zero-mode can be viewed as a homogeneous scalar field $\phi$ by its own \cite{hybr-inf4}. We call its canonical momentum $\pi_{\phi}$. Classically, and in absence of inhomogeneities, the system must satisfy a constraint which is equivalent to \begin{equation}
\label{clascon}
\pi_{\phi}^2-\frac{3}{4 \pi G \gamma^2} \Omega_0^2 + 8 \pi^2 G^2  \Delta \gamma^2 v^2  W(\phi)=0,
\end{equation} 
where we have used the notation $\Omega_0$ for $2\pi G \gamma bv$ and, for the time being, we have allowed the inclusion of a non-vanishing potential $W$ for the scalar field. 

In LQC, one frequently uses a representation in which the operator for $v$ (as well as that for $\phi$) acts by multiplication, with an inner product that is discrete in this volume variable \cite{LQC,AS,sigma}. The remarkable features of LQC, distinctive of this quantization in comparison with more standard ones, come precisely from this discreteness of the measure of integration over the volume. Moreover, physical states decouple in superselection sectors which are spanned by volume eigenstates that differ between them in multiples of four units \cite{LQC,onestep}. Now, for the superselection sector formed by all states with support on volumes equal to a multiple of four, namely $v=4n$ with $n$ an integer,\footnote{It is actually possible to see this sector as the union of two independent semilattices formed by positive and negative volumes, since the action of the geometric part of the constraint vanishes at zero volume \cite{sLQC,onestep}.} a especially manageable representation is obtained by performing a discrete Fourier transform from $v$ to the $b$ variable and implementing the following change:
\begin{equation}
\label{xslqc}
x=\frac{1}{\sqrt{12\pi G}} \ln{\Big[\tan{\Big(\frac{b}{2}\Big)}\Big]},
\end{equation}
so that $b=2 \tan^{-1}(e^{\sqrt{12\pi G} x})$. 

Representing the functions of the connection in terms of holonomy elements of the form $e^{\pm  i b/2}$, one obtains an operator counterpart of $\Omega_0^2$ which amounts to the replacement of $b$ with $\sin{b}$ \cite{sLQC}. Suppose then that we start with wave functions $\Gamma$ in the $(v,\phi)$-representation. If we introduce the scaling $\chi=\Gamma/(\pi v)$ and carry out the explained change to the $(x,\phi)$-representation, it turns out that the constraint of the homogeneous model, particularized to a vanishing field potential and with a densitization that can be associated with a harmonic time gauge (that simplifies considerably the factor ordering) \cite{sLQC}, adopts the expression
\begin{equation}
\label{SLQCcon}
\hat{\pi}_{\phi}^2-\hat{\pi}_{x}^2=0,
\end{equation}
where the two momentum operators act as derivatives: $\hat{\pi}_{\phi}=-i\partial_{\phi}$ and $\hat{\pi}_{x}=-i\partial_{x}$. Notice that, up to a constant factor, $\hat{\pi}_{x}$ is just an operator representation of $\Omega_0$:
\begin{equation}
\label{Omega0hat}
\hat{\Omega}_0 \equiv \sqrt{\frac{4\pi G \gamma^2}{3}} \hat{\pi}_x.
\end{equation}

The quantum dynamics generated by the constraint of sLQC is so simple that one can integrate it almost straightforwardly. Replacing the evolution time in terms of the scalar field $\phi$ one obtains that, whereas $\hat{\pi}_{\phi}$ and $\hat{\pi}_x$ are Dirac observables and hence preserved by the dynamics, the operator $\hat{x}$ satisfies
\begin{eqnarray}
\label{evolsLQC}
\hat{x}=\hat{x}_0+ (\phi-\phi_0) \,{\rm sign}(\hat{\pi}_x).
\end{eqnarray}
Here, $\hat{x}_0$ denotes the operator $\hat{x}$ on the section where the configuration of the scalar field (that serves as internal time) is $\phi_0$. We can understand $\phi=\phi_0$ as the initial section for the evolution. In the integration of $\hat{x}$, we have used that, in sLQC, one restricts the operator $\hat{\pi}_{\phi}$ to be positive in order to remove the double counting of solutions owing to time reversal invariance \cite{sLQC}.

Physical states have the form
\begin{equation}
\label{sLQCphys}
\chi(x,\phi)=\frac{1}{\sqrt{2}}[F(x_+)-F(x_-)],
\end{equation}
where $x_{\pm}=\phi\pm x$, corresponding to left and right moving modes, respectively, and $F$ is any function with Fourier transform supported on the positive real line. The inner product on physical states can be expressed, e.g. using only left moving modes, as
\begin{equation}
\label{sLQCinner}
(\chi_1,\chi_2)=-2i\int_{\mathbb R} dx F^*_1(x_+)\partial_xF_2(x_+).
\end{equation}
The symbol $*$ denotes complex conjugation. With this inner product, the operator $\hat{\pi}_x$ is positive on the sector of left moving modes, and negative for right moving modes. 

If one calls $\hat{P}_R$ and$\hat{P}_L$ the projectors on the right and left moving modes, it was shown in Ref. \cite{sLQC} that
\begin{equation}
\label{sLQCvol}\hat{v}=\frac{1}{\sqrt{3\pi G}}\sum_{j=R,L} \hat{P}_j \cosh{(\sqrt{12\pi G}\hat{x})} \hat{\pi}_x \hat{P}_j.
\end{equation}
Then, recalling the definition of the physical volume, it is straightforward to get its representation as the operator $\hat{V}=2\pi G \gamma  \sqrt{\Delta}|\hat{v}|$. 

Apart from the volume, we will need some other operators in order to include a potential for the scalar field and describe the interaction with the inhomogeneities later on in this work. Specifically, we will need operators for the regulated inverse volume and for the Hubble parameter. Besides, it will prove helpful to compute the explicit form of the operator\footnote{Let us comment that $\hat{\Omega}_0^2$ has a continuum spectrum \cite{onestep}; therefore the inverse that appears in our expression can be defined.} $\hat{B}=\sqrt{4 \pi G /3} \gamma \hat{V} |\hat{\Omega_0}|^{-1}\hat{V}$.  

Using a regularization of the inverse of the volume which is standard in LQC \cite{LQC,hybr-inf4} and expression \eqref{sLQCvol} for the volume operator, one can directly define 
\begin{equation}
\label{sLQCinverse}
\left[\widehat{\frac{1}{V}}\right]= \left(\frac{3}{2}\right)^3 \frac{1}{2\pi G \gamma\sqrt{\Delta}} \hat{v} \big( |\hat{v}+1|^{1/3}-|\hat{v}-1|^{1/3}\big)^3.
\end{equation}

As for the Hubble parameter, apart from factors of the volume, it can be expressed in terms of the analog of $\Omega_0$ when the length of the holonomies is doubled. This doubling is necessary if we want a linear operator on the superselection sector of sLQC \cite{hybr-inf4}, with support on volumes $v$ that are multiples of four units. In fact, our superselection sector is stable under $\hat{\Omega}_0^2$, and therefore under the homogeneous constraint, as well as under $|\hat{\Omega}_0|$, but not under the action of just $\hat{\Omega}_0$. Recalling that the effective counterpart of $\hat{\Omega}_0$ is $2\pi G \gamma \sin{b} \,v$, let us define 
\begin{equation}
\label{sLQCLambda}
\Lambda_0=2\pi G \gamma \sin{(2b)} \frac{v}{2}= 2\pi G \gamma \cos{b} \sin{b} \,v,
\end{equation} obtained by replacing the canonical set $\{v,b/2\}$ with the new set $\{v/2,b\}$, so that $b$ has half the period. Expressing the periodic functions above in terms of $v$ and (the effective) $\Omega_0$, and using their operator representations, a careful calculation (with a judicious choice of factor ordering) leads on each sector of left and right moving modes to 
\begin{equation}
\label{sLQChatLambda}
\hat{\Lambda}_0=-\sqrt{\frac{4\pi G\gamma^2}{3}} \tanh{(\sqrt{12\pi G} \hat{x})} \hat{\pi}_x .
\end{equation}

Finally, a straightforward computation shows that, on each of the sectors,
\begin{equation}
\label{sLQCtheta}
\hat{B}= \frac{4 \pi G \Delta \gamma^2}{3} \cosh^2{(\sqrt{12 \pi G} \hat{x})} |\hat{\pi}_x| .
\end{equation}

Let us conclude the section by noticing that, in sLQC, the quantum evolution of our auxiliary operators (the volume, its regulated inverse, $\hat{\Lambda}_0$, and $\hat{B}$) is that dictated by the evolution of $\hat{x}$ in Eq. \eqref{evolsLQC}, while $\hat{\pi}_x$ remains constant, as it corresponds to a Dirac observable.

\section{Perturbations and Effective Mukhanov-Sasaki equations}\label{sec:MS}

In the model described in the previous section, we now introduce a potential for the scalar field and study metric and field perturbations. The quantum treatment of this system was discussed in Refs. \cite{hybr-inf3,hybr-inf4}, truncating the action at quadratic order in the perturbations. For simplicity, we will consider only scalar perturbations, which are the relevant ones for present observations of the CMB. Tensor perturbations are easier to deal with, since they can be straightforwardly described by gauge invariants without mixing metric and field perturbations. Vector perturbations, on the other hand, do not contain any physical degree of freedom if the matter content is a scalar field.

The perturbations of the metric (including the induced spatial metric, the lapse function, and the shift) and the perturbations of the scalar field can be expanded in Fourier series and described by the time varying coefficients of this mode expansion (in this expansion, zero-modes are treated exactly at quadratic perturbative order, i.e., no perturbation of zero-modes is treated as independent at that order \cite{hybr-inf4}). The introduced inhomogeneities are subject to two types of constraints that arise, respectively, from the perturbation of the Hamiltonian and the momentum constraints of General Relativity around the FLRW model with field potential. Although these perturbative constraints do not commute, it is possible to  find an Abelianization of them at the order of truncation adopted in the action. The MS gauge invariants commute with these perturbative constraints (as well as with their Abelianized version). It is then possible to find a complete set of canonical variables for the perturbations formed by the MS gauge invariants, the MS momentum variables (which are also gauge invariants), the Abelianized perturbative constraints, and canonical momenta for these constraints. Actually, the MS momenta can be uniquely determined, removing the ambiguity in adding a contribution linear in the MS configuration variables, by demanding that one can find a Fock quantization for the MS field such that the vacuum is invariant under the isometries of the spatial sections and the MS dynamics admits a unitary implementation \cite{unitary}. Once this canonical set for the perturbations has been constructed, it is possible to complete it into a canonical set for the entire system, including zero-modes. The original zero-modes must be corrected with some fixed, quadratic perturbative contributions. The new zero-modes are those that must be identified with the variables of the model discussed in Sec. \ref{sec:sLQC}. The final result of the reformulation of the system is a full description designed to quantize the system without any need to fix the perturbative gauge and prepared to extract physics using gauge invariants for the inhomogeneities \cite{hybr-inf4}.

Under quantization, physical states are independent of the variables conjugate to the Abelianized perturbative constraints. Hence, physical states depend only on the MS gauge invariants and on the zero-modes of the FLRW geometry and the scalar field. Following the notation of Ref. \cite{hybr-inf4}, we will call $v_{\vec{n},\epsilon}$ the (real) Fourier coefficients of the MS field, where $\vec{n}$ is the wave vector of the Fourier mode and $\epsilon=\pm$ depending on whether it is a sine or a cosine mode. In more detail, $\vec{n}$ is a triple of integers, owing to the periodicity of the coordinates on $T^3$, and its first non-vanishing component is chosen to be positive, in order to avoid a double counting of modes, since each wave vector is already associated to a sine and a cosine Fourier component.

The entire system is still subject to a constraint that is non-trivial to impose: The zero-mode of the Hamiltonian constraint. This constraint equals the corresponding constraint of the homogeneous and isotropic FLRW model plus a quadratic contribution of the perturbations, which incorporates backreaction effects at the order of truncation adopted in the action. Using the prescriptions detailed in Ref. \cite{hybr-inf4}, and up to an irrelevant numerical global factor, this Hamiltonian constraint can be represented by the operator $\hat {\mathcal C_T}=\hat {\mathcal C}^{(0)}+\sum_{\vec n,\epsilon}{\hat{\mathcal C}}^{\vec n,\epsilon}$, where the contribution of the homogeneous sector, including a field potential $W(\phi)$, is 
\begin{eqnarray}\label{eq:calC_0}
\hat{\mathcal C}^{(0)} &=& {\hat \pi}_\phi^2- \hat{\mathcal H}_0^{(2)},
\\
\label{eq:calH_0}
\hat{\mathcal H}_0^{(2)} &=&\frac{3\hat\Omega_0^2}{4\pi G \gamma^2}-2 W(\hat\phi)\hat{V}^2=\hat{\pi}_x^2- 2 W(\hat\phi)\hat{V}^2.
\end{eqnarray}
Although we adopt here the operator  $2W(\hat\phi)\hat{V}^2$ as a natural choice to represent the contribution of the potential, we would like to leave open the possibility of other representations, corresponding  to different factor orderings. We will return to this point later on, in Sec. \ref{sec:potential}.

On the other hand, the quadratic contributions of the perturbations are
\begin{equation}
{\hat{\mathcal C}}^{\vec n,\epsilon}= - \frac{4\pi G}{3}\left[\hat{\Theta}^{\vec n,\epsilon}_{e}+ \big(\hat{\Theta}^{\vec n,\epsilon}_{o}{\hat \pi}_{\phi}\big)_{sym}\right],
\end{equation}
where the subindex of the parentheses stands for symmetrization of the product of operators, and we have called
\begin{eqnarray}\label{eq:perturbation-operators}
\hat{\Theta}_o^{\vec n,\epsilon} &=&  - \hat{\vartheta}_o \hat{v}_{\vec n,\epsilon}^2 \\
\hat{\Theta}_e^{\vec n,\epsilon} &=& -\left[(\hat{\vartheta}_e \omega_n^2+\hat{\vartheta}_e^q)\hat{v}_{\vec n,\epsilon}^2+ \hat{\vartheta}_e \hat{\pi}_{v_{\vec n,\epsilon}}^2\right].
\end{eqnarray}
Here, the square frequency of the mode is the square norm of its wave vector, $\omega_n^2=\vec{n} \cdot\vec{n}$, and (taking into account Eq. (6.6) in Ref. \cite{hybr-inf4} and our convention for numerical factors in this work) we have introduced the following functions of the zero-modes:
\begin{eqnarray}
\hat\vartheta_o&=& 4\sqrt{\frac{3G}{\pi} }\gamma W^{\prime}(\hat\phi) \hat V^{2/3} |\hat\Omega_0|^{-1} \hat\Lambda_0|\hat\Omega_0|^{-1}\hat V^{2/3} ,\\
\hat\vartheta_e&=&\frac{3 }{2 G}\hat V^{2/3},\\
\hat\vartheta_e^q&=&\frac{1}{2\pi}\widehat{\left[\frac1{V}\right]}^{1/3}\hat{\mathcal H}_0^{(2)}\left(19-24 \pi G \gamma^2 \hat\Omega_0^{-2}\hat{\mathcal H}_0^{(2)}\right) \widehat{\left[\frac1{V}\right]}^{1/3} \nonumber\\
& +&\frac{3 }{8 \pi^2 G  }\hat V^{4/3}\left( W^{\prime\prime}(\hat{\phi})- \frac{16\pi G}{3} W(\hat\phi)\right).
\end{eqnarray}
The prime stands for the derivative with respect to $\phi$. 

Notice that all the operators in these expressions are  available in sLQC according to our discussion in the previous section. The inverse operators $\hat\Omega_0^{-2}$ and $|\hat\Omega_0|^{-1}$ can be constructed from the positive operator $\hat\Omega_0^{2}$ (and its square root, that gives $|\hat\Omega_0|$) using the spectral theorem.

Of particular interest are states that satisfy an ansatz similar to the Born-Oppenheimer ansatz of molecular and atomic physics, in the sense that their dependence on the perturbations and on the zero-mode of the FLRW geometry can be separated. This ansatz amounts to a factorization of the wave function into two factors, namely, $\Psi=\chi(x,\phi) \psi({\mathcal N},\phi)$. Here, the argument $x$ of the wave function $\chi$ denotes simply dependence on the zero-mode of the FLRW geometry, while the dependence on the MS field is indicated with the label ${\mathcal N}$ which represents the set of occupancy numbers of the MS modes in a privileged Fock quantization selected, up to unitary equivalence, by the criteria of invariance under the spatial isometries and unitary evolution (in the regime of deparametrized dynamics) \cite{unitary}. Furthermore, we assume a state $\chi$ with a unitary dynamics in its $\phi$-dependence. The dynamics will be generated by an operator $\hat{\mathcal H}_0$ that we take to coincide with the square root of $\hat{\mathcal H}_0^{(2)}$, at least up to terms that, when squared, are negligible compared to the quadratic perturbative contributions of the constraint.\footnote{ If the original $\hat{\mathcal H}_0^{(2)}$ is not positive but just self-adjoint, in practice, for small perturbations and given Eq. \eqref{eq:calC_0}, it suffices to consider the positive part of its spectrum taking projections, and proceed then to calculate the square root.} The operator $\hat{\mathcal H}_0$ can be regarded as a modification of the evolution generator (along $\phi$) of the homogeneous system with a free massless field in order to include a field potential. In summary, we adopt for $\chi$ the ansatz $\chi(x,\phi)=\hat{U}(x,\phi)\chi_0(x)$, where $\chi_0$ is the initial FLRW state at a certain value $\phi_0$ of the homogeneous scalar field, and
\begin{align}\label{chievolution}
\hat U(x;\phi)={\mathcal P}\left[\exp{\left(i\int^{\phi}_{\phi_0} d{\tilde{\phi}}\,\hat{\mathcal H}_0(x,\tilde{\phi})\right)}\right].
\end{align}
The symbol $\mathcal P$ denotes time ordering with respect to $\phi$. 

We will specialize our analysis to the $\chi$-representation of sLQC, although the discussion can be carried out in other, more general quantizations. 

The dynamics of the perturbations obtained when this ansatz is substituted in the constraint were discussed in Ref. \cite{hybr-inf4}. If one disregards the possible quantum transitions between different states of the homogeneous FLRW geometry mediated by the action of the constraint, one gets, without further assumptions, a relatively simple evolution of the inhomogeneities. The master equation for this evolution can be interpreted as a constraint on $\psi$ given by
\begin{eqnarray}
\label{constperturb}
&&\hat{\pi}_{\phi}^2+ 2 \langle \hat{{\mathcal H}}_0 \rangle_{\chi}  \hat{\pi}_{\phi} - E_{\chi}(\phi) \nonumber\\
&&- \langle  \hat{\Theta}_{e} + (\hat{\Theta}_{o}\hat{{\mathcal H}}_0)_{sym}+\frac{1}{2}[ \hat{\pi}_{\phi}-\hat{\mathcal H}_0, \hat{\Theta}_{o}] \rangle_{\chi}=0,
\end{eqnarray}
where $ \hat{\Theta}_{p}=\sum_{\vec{n},\epsilon} \hat{\Theta}_{p}^{\vec{n},\epsilon}$ for $p=e,o$ and  $E_{\chi}$ is a state-dependent function of the homogeneous scalar field. This expression follows from Eqs. (5.12) and (5.15) of Ref. \cite{hybr-inf4} with our convention of numerical factors and removing a negligible perturbative  correction to the second term, linear in $\hat{\pi}_{\phi}$.\footnote{The commutator in this constraint is usually supposed ignorable, and can even be regarded as a term absorbable with a different factor ordering in $\sum_{\vec{n},\epsilon} \hat{\cal C}^{\vec{n},\epsilon}$; though, we will carry on our analysis with it. Besides, $E_{\chi}$ is taken at most of the perturbative order of the $\Theta$-terms, adapting the factor ordering for this if necessary.} The expectation values are taken on the state $\chi$ with respect to the zero-mode of the FLRW geometry: In our case with the inner product of sLQC. 

Admitting that the quadratic dependence on operators acting on the perturbations has a direct translation into the same dependence on classical variables, the above constraint on the inhomogeneities has an effective counterpart which leads to  modified MS equations. Explicitly, one obtains the mode equations
\begin{eqnarray}
d^2_{\eta_{\chi}}v_{\vec n,\epsilon} &=&
- \frac{  \langle \hat{\vartheta}_{e}^q + (\hat{\vartheta}_{o}\hat{{\mathcal H}}_0)_{sym}+\frac{1}{2}[ \hat{\pi}_{\phi}-\hat{\mathcal H}_0, \hat{\vartheta}_{o}]\rangle_{\chi}}{ \langle \hat{\vartheta}_{e}\rangle_{\chi}} v_{\vec n,\epsilon} \nonumber \\
&-&  \omega_n^2  v_{\vec n,\epsilon} . \label{MSeqhybrid}
\end{eqnarray}
The time appearing in these effective equations is a conformal time given by $d\eta_{\chi}=\langle\hat{\vartheta}_e\rangle_{\chi} dT$, where $T$ is the harmonic time\footnote{Actually, $T$ is the time parameter of the evolution generated by half the effective constraint obtained from Eq. \eqref{constperturb} (see Ref. \cite{hybr-inf4}).} adopted in Sec. \ref{sec:sLQC}. Similarly, from the effective constraint on the perturbations and the definition of $T $, one gets the relation $d\phi= (\pi_{\phi}^{{\rm (inh)}}+\langle \hat{\cal H}_0\rangle_{\chi}) dT$. Here, we have used the superscript ``${\rm (inh)}$'' to emphasize that this term provides just the contribution of the inhomogeneous sector to the zero-mode of the scalar field momentum. At the order of the mode equations \eqref{MSeqhybrid}, we expect that this contribution should be ignorable compared with $\langle\hat{\cal H}_0\rangle_{\chi}$ (assuming that the latter is not negligibly small), because $\pi_{\phi}^{{\rm (inh)}}$ should be of perturbative order according to our effective constraint. Combining this result with the definition of the conformal time, we conclude
\begin{equation} \label{phiconftime}
\langle \hat{\cal H}_0\rangle_{\chi} d\eta_{\chi} = \langle\hat{ \vartheta}_e\rangle_{\chi}  d\phi.
\end{equation}
It is worth remarking that this change to our conformal time is state dependent.

The ratio of expectation values in Eq. (\ref{MSeqhybrid}) contains quantum modifications with respect to the corresponding term in the standard MS equations. Therefore, it becomes clear that the calculation of these expectation values is a central issue to discuss the possible quantum effects that affect the power spectrum of the CMB.   

\section{Interaction picture}\label{sec:Interaction}

The main problem to calculate the required expectation values is the integration of the evolution of the FLRW state $\chi$, provided by ${\hat{\mathcal H}}_0$, when the scalar field potential is not constant, since the dynamics is not solvable then and there exist complications even for the numerical integration. The problem can be alleviated by extracting the dynamics of the free-field case, with a vanishing potential, treating the rest of the evolution as a kind of geometric interaction and passing,  consequently, to an interaction picture \cite{GalindoPascual}, in which the homogeneous field plays the role of evolution time. We will implement this idea in the present section.

Let us consider the expectation value $\langle \hat{A}(\phi) \rangle_{\chi}$  of a generic operator $\hat{A}(\phi)$ on the state $\chi(\phi)$ of the FLRW geometry, where we are showing explicitly the possible dependence on the variable $\phi$. We first define the operator ${\hat{\mathcal H}}_0$ for vanishing potential:
\begin{equation} \label{freeham}
{\hat{\mathcal H}}_0^{(F)}=\sqrt{\frac{3}{4 \pi G \gamma^2 }} |\hat{\Omega}_0|.
\end{equation}
Notice that this operator is independent of $\phi$. Besides, we have ${\hat{\mathcal H}}_0^{(F)}=|\hat{\pi}_x|$ (e.g. in sLQC). We can now introduce the counterpart of the state $\chi$ in the interaction picture,
\begin{equation}
\label{intstate}
\chi_I(x,\phi)=e^{-i{\hat{\mathcal H}}_0^{(F)}(\phi-\phi_0)}\chi(x,\phi),
\end{equation}
where $\phi_0$ is the initial value of $\phi$. For any operator $\hat{A}$ in the original Schr\"odinger-like picture, the corresponding operator in the interaction picture is given by \cite{GalindoPascual}
\begin{equation}
\label{operatorint}
\hat{A}_I= e^{-i{\hat{\mathcal H}}_0^{(F)}(\phi-\phi_0)} \hat{A} e^{i{\hat{\mathcal H}}_0^{(F)}(\phi-\phi_0)}.
\end{equation}

Then, if we call ${\hat{\mathcal H}}_1={\hat{\mathcal H}}_0-{\hat{\mathcal H}}_0^{(F)}$, it is a well known (and easily reproducible) result that the evolution of $\chi_I$ is generated by the operator ${\hat{\mathcal H}}_{1I}$. Therefore
\begin{eqnarray}\label{Uint}
\chi_I(x,\phi)&=& \hat{U}_I(x,\phi)\chi_0(x),\\ \label{Uopint}
\hat{U}_I(x,\phi)&=& {\mathcal P}\left[\exp{\left(i\int^{\phi}_{\phi_0} d{\tilde{\phi}}\,\hat{\mathcal H}_{1I}(x,\tilde{\phi}) \right)}\right].
\end{eqnarray}
In total, we find that
\begin{equation}\label{expvalueint}
\langle \hat{A}(\phi) \rangle_{\chi}= \langle \hat{A}_I(\phi) \rangle_{\chi_I}=
\langle \hat{U}_I^{\dagger}(\phi)  \hat{A}_I(\phi) \hat{U}_I(\phi) \rangle_{\chi_0},
\end{equation}
where the dagger denotes the adjoint.

In Sec. \ref{sec:sLQC}, we saw that the integration of the dynamics of the free case can be performed analytically in sLQC. In that case, the form of the FLRW-geometry operators in the interaction picture is in fact straightforward to obtain. It suffices to replace their dependence on the basic operator $\hat{x}$ by the same dependence on the evolved operator according to Eq. \eqref{evolsLQC}, namely
\begin{equation}\label{freeevol}
\hat{x} \rightarrow \hat{x}(\phi)=\hat{x} + (\phi-\phi_0)\,{\rm sign}(\hat{\pi}_x) .
\end{equation} 
This reduces the problem of the dynamical evolution of our expectation values to the computation of the path-ordered integral in the definition \eqref{Uopint}.

\section{Quantum contributions of the potential}\label{sec:potential}

The quantum dynamics generated by $\hat{\mathcal H}_{1I}$ is still too complicated to be manageable in practice  when a field potential is present. One may treat this evolution semiclassically, assuming that the state $\chi$ displays a semiclassical behavior. Notice that, in principle, this concept of semiclassicality is assigned now to the trajectories generated by the interaction evolution, since the free-field contributions were already accounted for with our treatment in the previous section. In order not to restrict ourselves necessarily to this semiclassical regime, in this section we will go further in our analysis and extract the dominant contributions of the potential in the quantum evolution. To simplify the notation, we admit from now on that the representation of the homogeneous scalar field acts as a multiplicative operator. 

Let us study the operator $\hat{\mathcal H}_{1I}$ in more detail. We recall that this operator is the translation into the interaction picture of the difference $\hat{\mathcal H}_{1}$ between the evolution generator in the homogeneous case (the square root of $\hat{\mathcal H}_{0}^{(2)}$) and its counterpart $\hat{\mathcal H}_0^{(F)}$ in the free-field scenario with vanishing potential $W(\phi)$. The passage to the interaction picture is done straightforwardly by implementing the replacement \eqref{freeevol}. We now want to show that, up to cubic terms and higher orders in the potential, the operator $\hat{\mathcal H}_{1}$ can be represented with a suitable choice of factor ordering in the following approximate form:
\begin{eqnarray}
\hat{\mathcal H}_{1} &\approx& \hat{\mathcal H}_{2}=-  W(\phi) \hat{B} 
- W^2(\phi)\hat{C} , \label{dominantint} \\
\hat{C} &= &   \sqrt{\frac{\pi G}{3}} \gamma |\hat{\Omega}_0|^{-1/2}\,\hat{B}^2 \,|\hat{\Omega}_0|^{-1/2}, \label{Boperator}
\end{eqnarray}
where we have defined a new geometric operator $\hat{C}$ that provides the part quadratic in the potential, and employed the operator $\hat{B}=\sqrt{4 \pi G/3}  \gamma \hat{V} |\hat{\Omega}_0|^{-1} \hat{V}$ introduced in Sec. \ref{sec:sLQC}. 

It will suffice to prove that the square of $\hat{\mathcal H}_0^{(F)}+\hat{\mathcal H}_{2}$ is an acceptable representation of $\hat{\mathcal H}_{0}^{(2)}$ at the desired order. A careful but nonetheless direct calculation leads to
\begin{eqnarray}
&&\left(\sqrt{\frac{3}{4\pi G \gamma^2}} |\hat{\Omega}_0|+\hat{\mathcal H}_{2}\right)^2= \frac{3}{4\pi G \gamma^2} \hat{\Omega}_0^2 - 2 W(\phi) \hat{V}^2 \nonumber \\
&&- W(\phi) \hat{Q}_1  -  W^2(\phi) \hat{Q}_2+\mathcal{O}(W^3),
\label{square}
\end{eqnarray}
where we have introduced two operators $\hat{Q}_1$ and $\hat{Q}_2$ that are independent of the scalar field, and hence act only on the zero-mode of the FLRW geometry. More importantly, they are pure commutators and so they can be interpreted as arising from a particular choice of factor ordering. Furthermore, in fact they are double commutators, and therefore they provide second-order quantum corrections. Explicitly, these operators are
\begin{eqnarray}
\hat{Q}_1&=& \big[|\hat{\Omega}_0|^{-1},\hat{V}\big]\big[|\hat{\Omega}_0|,\hat{V}\big]+\big[ \big[|\hat{\Omega}_0|,\hat{V}\big],|\hat{\Omega}_0|^{-1}\hat{V}\big],\nonumber\\
\hat{Q}_2&=& \frac{1}{2}\big[ \big[|\hat{\Omega}_0|^{1/2},\hat{B}^2\big],|\hat{\Omega}_0|^{-1/2}\big] .
\label{Q}
\end{eqnarray}

In total, if we neglect terms that are cubic or of higher order in the field potential, we conclude that the operator \eqref{square} is a viable representation of the generator of the homogeneous evolution. Thus, we can write
\begin{eqnarray}
\label{split}
\hat{\mathcal H}_{1I} = \hat{\mathcal H}_{2I} + \hat{\mathcal H}_{3I}, 
\end{eqnarray}
where $\hat{\mathcal H}_{2I} $ is the operator \eqref{dominantint} in the interaction picture, and $\hat{\mathcal H}_{3I}$ is just the remaining part of the original interaction operator, which is at least of cubic order in the potential. 

Obviously, if we remove the quadratic contribution of $W$ in $\hat{\mathcal H}_{2}$ [given by the last term in Eq. \eqref{dominantint}], our formula \eqref{split} continues to be valid, though then $\hat{\mathcal H}_{3I}$ would be of quadratic order in the field potential. The convenience of keeping or eliminating in $\hat{\mathcal H}_{2}$ terms quadratic in $W$ depends on how relevant the quantum contribution of those terms turns out to be in order to obtain accurate results.

In the case of sLQC, recalling that $\hat{\Omega}_0$ is proportional to $\hat{\pi}_x$ and hence it is a Dirac observable for the free field, the dominant contribution of the potential to the generator of the evolution in the interaction picture can be obtained from Eqs. \eqref{dominantint} and \eqref{Boperator} just by performing the substitution \eqref{freeevol} in the $x$-dependence of the operator $\hat{B}$ [namely, in the square hyperbolic cosine of Eq. \eqref{sLQCtheta}].

We can now extract the dynamics generated by $\hat{\mathcal H}_{2I}$, following the same steps as if we introduced a new interaction image.
If we call
\begin{equation}\label{U2I}
{\hat U}_{2I} = {\mathcal P} \left[ \exp{ \left( i \int^{\phi}_{\phi_0} d{\tilde{\phi}} \hat{\mathcal H}_{2I} (\tilde{\phi}) \right) } \right],
\end{equation}
and, for any operator $\hat{A}_I$ in the original interaction picture, we define
\begin{equation}\label{2intpicture}
\hat{A}_{J}= \hat{U}_{2I}^{\dagger}\hat{A}_I \hat{U}_{2I},
\end{equation}
then the expectation value of the operator is given by
\begin{equation}
\label{expvalues2int}
\langle \hat{A}_I \rangle_{\chi_I}= \langle \hat{U}_{J}^{\dagger} \hat{A}_{J} \hat{U}_{J}\rangle_{\chi_0},
\end{equation}
where 
\begin{equation}\label{UII}
\hat{U}_{J}= {\mathcal P}\left[\exp{\left(i\int^{\phi}_{\phi_0} d{\tilde{\phi}}\,\hat{\mathcal H}_{3J} (\tilde{\phi})\right)}\right].
\end{equation}

So far, our treatment of the expectation values is exact. The obstruction that we find now is the integration of the evolution generated by $\hat{\mathcal H}_{2I}$ and by $\hat{\mathcal H}_{3J}$ in order to calculate the unitary operators $\hat{U}_{2I}$ and $\hat{U}_J$, respectively. Concerning $\hat{\mathcal H}_{2I}$, one can try to compute the associated evolution by determining its eigenfunctions numerically. In particular, if one does not include quadratic contributions of the potential in $\hat{\mathcal H}_{2I}$, its expression looks manageable enough as to allow for the calculation of its spectrum. Another possibility, obviously, is to renounce to the exact treatment at this stage and introduce approximations. In particular, we can truncate the series expansion of ${\hat U}_{2I} $ in terms of path ordered integrals of powers of $\hat{\mathcal H}_{2I}$ \cite{GalindoPascual} so as to compute the operator $\hat{A}_J$ up to a certain order of the potential. Obviously, the unitarity of the change of representation is broken in this truncation, but just at the order that is neglected. In the case that the potential is a mass term, the approximation can be regarded as a truncation of the asymptotic expansion in powers of the squared mass. For the evolution operator $\hat{U}_J$, on the other hand, an appealing proposal consists in treating these dynamics in an effective approximation. 

At linear order in the potential, Eq. \eqref{2intpicture} can be approximated as
\begin{equation}
\label{linear2intpicture}
\hat{A}_J\approx \hat{A}_I -i \Big[\hat{A}_I,\int^{\phi}_{\phi_0}d{\tilde \phi} W({\tilde \phi})
\hat{B}_I({\tilde \phi}) \Big].
\end{equation}
The notation $\hat{B}_I(\phi)$ indicates that the operator $\hat{B}$ in the considered interaction picture depends on $\phi$, since it equals its evolution along this internal time parameter for the free-field case, i.e., when the potential vanishes. 

Moreover, if one wants to keep quadratic contributions of the potential in the operator $\hat{A}_J$, these are given by the additional terms
\begin{eqnarray}
&-  &  \hat{A}_I  \int^{\phi}_{\phi_0} d{\tilde \phi} W(\tilde \phi)\hat{B}_I(\tilde \phi) \int^{{\tilde \phi}}_{\phi_0}  d{\bar \phi} W(\bar \phi) \hat{B}_I(\bar \phi)  \nonumber \\ &-  & \Big(\int^{\phi}_{\phi_0}d{\tilde \phi} W(\tilde \phi)\hat{B}_I(\tilde \phi)\int^{{\tilde \phi}}_{\phi_0} d{\bar \phi} W(\bar \phi) \hat{B}_I(\bar \phi) \Big) \hat{A}_I \nonumber \\ &+&\Big(\int^{\phi}_{\phi_0} d{\tilde \phi} W(\tilde \phi) \hat{B}_I(\tilde \phi) \Big) \hat{A}_I \Big( \int^{\phi}_{\phi_0} d{\tilde \phi} W(\tilde \phi) \hat{B}_I(\tilde \phi) \Big) \nonumber \\ &- i & \Big[ \hat{A}_I, \int^{\phi}_{\phi_0} d{\tilde \phi} W^2(\tilde \phi) \hat{C}_I(\tilde \phi) \Big],
\label{quadraticAJ}
\end{eqnarray}
where $\hat{C}_I$ is the operator \eqref{Boperator} in our interaction picture.

Let us now particularize our discussion to the FLRW quantization given in sLQC and to a field potential equal to a mass term, $W(\phi)= m^2 \phi^2/2$. We recall that, in sLQC, the operator $\hat{B}_I$ is obtained from $\hat{B}$ by replacing its dependence on $\hat{x}$ with the same dependence on $\hat{x}(\phi)$, defined in Eq. \eqref{freeevol}. A careful calculation shows then that
\begin{equation}
\label{Bintegrated}
\int^{\phi}_{\phi_0}d{\tilde \phi} W({\tilde \phi})\hat{B}_I({\tilde \phi})=\frac{2\pi G}{3}m^2\Delta\gamma^2 \left( \hat{F}(\phi)-\hat{F}(\phi_0)\right) \hat{\pi}_x, 
\end{equation}
where we have defined the operator
\begin{eqnarray}
\hat{F}(\phi)& = & \frac{1}{8\sqrt{ 3\pi G}}\bigg(\phi^2 + \frac{1}{24\pi G} \bigg)\sinh{\bigg(4\sqrt{3\pi G} \hat{x}(\phi)\bigg)}  \nonumber\\
& -& \frac{\phi}{48\pi G}\cosh{\bigg(4\sqrt{3\pi G}\hat{x}(\phi)\bigg)}{\rm sign}(\hat{\pi}_x)\nonumber\\
&+&  \frac{\phi^3}{6} {\rm sign}(\hat{\pi}_x).\label{Foperator}
\end{eqnarray}
The quadratic contributions \eqref{quadraticAJ} for the massive field in sLQC are computed explicitly in the Appendix.

\section{Remaining dynamics}\label{sec:Remaining}

In the previous section we split the interaction generator $\hat{\cal H}_{1I}$ in two terms: A contribution containing the first powers of the potential, $\hat{\cal H}_{2I}$, and a remnant that we called $\hat{\cal H}_{3I}$.  We then truncated the evolution operator corresponding to $\hat{\cal H}_{2I}$ at lowest orders in the potential. 

An inspection of Eqs. \eqref{linear2intpicture} and \eqref{quadraticAJ} leads us to  expect that the terms neglected in the evolution operator $\hat{U}_{2I}$ are of the following relative order with respect to the conserved ones. For contributions of the operator $\hat{B}$, the relative order is 
\begin{equation}
R_B=\sqrt{G} \gamma |(\phi - \phi_0) W| \frac{V_I^2 }{|\Omega_0|},
\end{equation} 
where we use that $\Omega_0$ is a constant of motion in the absence of potential, and we {\emph{assume}} that the change of $WV_I^2$ is not too important in the interval $(\phi_0,\phi) $ (if this assumption is not valid, the estimation of the relative order will be more complicated). Therefore, the truncation should be valid if $R_B\ll 1 $. If quadratic contributions of the potential had not been taken into account in $\hat{\cal H}_{2I}$, this is the only restriction on the approximation. However, if such quadratic terms have been considered, hence including the operator $\hat{C}_I$ in the analysis, then the corresponding contributions in the expansion of the evolution operator are expected to be, for similar reasons as above, of a relative order 
\begin{equation}\label{RBC}
R_C=R_B^2 r_{BC}, \qquad r_{BC}= \frac{\sqrt{G}\gamma}{|\Omega_0 (\phi-\phi_0)|} .
\end{equation} 
In this case, the validity of the truncation requires that $R_C\ll 1$ as well.

When the conditions for the truncation cease to apply, one must either attempt a quantum integration of the whole evolution operator $\hat{U}_{2I}$, or renounce to extract a dominant contribution from the interaction generator $\hat{\cal H}_{1I}$ and explore instead, for instance, whether the whole interaction dynamics can be treated effectively. In the following, we focus the discussion on the region where the truncation is meaningful,  and discuss the remaining evolution operator $\hat{U}_J$, introduced in Eq. \eqref{UII}.

The lowest non-zero power of the potential that contributes to this remnant of the evolution gives a term of order $R_C$ if the interaction $\hat{\cal H}_{2I}$ does not contain quadratic factors of $W$ (because then $\hat{C}$ is included in $\hat{U}_J$, rather than in $\hat{\cal H}_{2I}$). This term is relevant in the dynamics, and hence also the evolution dictated by  $\hat{U}_J$, if its contribution is larger than the terms neglected in the truncation of ${\hat U}_{2I}$, which are of order $R_B^2$ or $R_B^3$, depending on whether the truncation keeps terms linear or square in $\hat{B}$, respectively. Since $R_B\ll 1$ for the validity of the truncation, we take e.g. the most stringent condition, $R_C\succ R_B^2$, where the symbol $\succ$ must be interpreted as the statement that the  factor on the left hand side is large compared to the other. Using our notation in Eq. \eqref{RBC}, this condition is equivalent to $r_{BC}\succ 1$. So, in this case where the operator $\hat{\cal H}_{2I}$ is defined without the inclusion of $\hat{C}$, the truncation of $\hat{U}_{2I}$ and the consideration of the remaining dynamics $\hat{U}_J$ is meaningful in the sector with $R_B\ll 1$ and $r_{BC}\succ 1$ (notice that this sector is not empty). Besides, the remaining evolution under discussion should not be ignored at least when the analyzed contribution of its lowest power in the potential cannot be neglected, namely, when $R_C$ is not small. Since the quantum treatment of the evolution  $\hat{U}_J$ is too intricate, the study has to be done in this case by means of an effective approximation. For this, it is necessary that the considered state of the FLRW geometry is peaked around an effective trajectory of the evolution generated by $\hat{\cal H}_{3J}$. 

For completeness, let us compile the expression of this generator, following all the steps that have been explained above, and adopting a truncation of $\hat{U}_{2I}$  at linear order in the potential:
\begin{eqnarray}
\label{H3Jexpress}
{\cal H}_{3J}&=& {\cal H}_{3I}+ \bigg\{{\cal H}_{3I}, \int^{\phi}_{\phi_0} d{\tilde \phi} W({\tilde \phi})B_I({\tilde \phi})\bigg\} \\
{\cal H}_{3I}&=& \left( \frac{3 \Omega_0^2}{4\pi G \gamma^2} - 2 W(\phi) V^2_I \right)^{1/2} -\sqrt{ \frac{3 }{4\pi G}} \frac{|\Omega_0|}{ \gamma}\nonumber\\
&+&  \sqrt{\frac { 4\pi G }{3}} \gamma W(\phi)\frac{V_I^2 }{|\Omega_0|},
\end{eqnarray}
where the curly brackets denote Poisson brackets, we call $B_I=\sqrt{4\pi G/3}\gamma V_I^2 /|\Omega_0|$, and the subindex $I$ in $V_I$ stands for the substitution of $x$ by $x+  (\phi-\phi_0)\, {\rm sign}(\pi_x)$ in the expression of the physical volume. The argument of the square root is assumed to be non-negative, otherwise the evolution must be set equal to zero.

If one has taken into account the quadratic contribution of the potential that goes with the operator $\hat{C}$ in the part $\hat{\cal H}_{2I}$ of the interaction, then, using the definition of the generator of the free homogeneous evolution $\hat{\cal H}_0^{(F)}$ and expression \eqref{dominantint}, one can convince oneself that the first contribution in powers of the potential in $\hat{U}_J$ is of order $R_B^3 r_{BC}^2$, while the terms ignored in the truncation of $\hat{U}_{2I}$ are expected to be of order $R_B^3$ and 
$R_B^3r_{BC}$. Therefore, the remaining interaction dynamics is relevant at the order of truncation provided that $r_{BC}\succ 1$, just as above when $\hat{\cal H}_{2I}$ included only factors that are linear in the potential. In the present case, the conditions for the truncation and the consideration of the remaining dynamics can be combined as follows: $1\gg R_C=R_B^2r_{BC}\succ R_B^2$ (where, we recall, the equality is a definition). In addition, the remaining part of the evolution, $\hat{U}_J$, should not be ignored now when $R_C^2/R_B$ is not much smaller than the unit. Similar comments as above apply to its possible approximation by means of an effective description, although now the expression of the generator ${\cal H}_{3J}$ is more complicated (but nonetheless straightforward to obtain).

\section{Summary and discussion}\label{sec:Conclu}

In quantum cosmology (including but not exclusively in LQC), the  MS equations that govern the behavior of the primordial fluctuations are modified with quantum corrections. The equations of the perturbation modes have coefficients that are determined by expectation values of quantum operators on the state that describes the FLRW geometry of  the Universe. These expectation values encode quantum phenomena of the geometry, and their expressions vary depending on the quantum approach adopted and on the specific quantization prescription followed within that approach. The possibilities to discriminate between the different  proposals seem to require analyses that retain the genuine quantum nature of the calculations, going beyond semiclassical or effective approximations which, though capturing the most relevant modifications to the classical description, ignore other quantum effects that may relevant for such discernment. The final hope is that confrontation of predictions with cosmological observations (mainly of the CMB) may eventually falsify the theoretical models. 

In this work we have proposed a number of steps aimed at facilitating the computation of those expectations values, by permitting some approximations while still retaining a significant part of their quantum features. The expectation values need to be calculated during the evolution of the FLRW state along the variation of the homogeneous part of the scalar field (namely, its zero-mode), that can be regarded as an internal time. This evolution is generated by an operator that differs from that of the FLRW geometry  with a free massless scalar field by contributions of the field potential. Therefore, the fist step in our treatment consists in extracting the free massless field part from the dynamics. We have done this by adopting an interaction picture. 

In a second step, one has to deal with the integration of the free-field dynamics, something that can be achieved analytically in certain approaches to quantum cosmology, like in a prescription for LQC known as sLQC. We have specialized our study to this quantization of the FLRW geometry in order to obtain manageable formulas in a concrete case, though the analysis can be performed in other cases or with other methods, for instance numerically (resolving the spectral decomposition associated with the evolution operator by means of numerical methods).

The dynamics remaining in the expectation values is generated by an interaction operator that vanishes when so does the field potential $W$. In a third step we have split this operator in two parts, one of them (called $\hat{\cal H}_{2I}$) capturing the contributions with the lowest powers of the potential. We have then passed to a new interaction image. The next step is the integration of the dynamics generated by $\hat{\cal H}_{2I}$. Now the problem is more intricate because, generically, an analytical integration is not known in any quantization scheme. Different avenues appear. 

One of them is to try and integrate the dynamics numerically. If only terms linear in the potential are kept in $\hat{\cal H}_{2I}$, this amounts to resolve the spectral decomposition associated with the operator $\hat{B}$ that, up to a numerical factor, equals the symmetric product $\hat{V}|\hat{\Omega}_0|^{-1}\hat{V}$ (notice that the resolution of the identity  of its interaction  image counterpart, denoted $\hat{B}_I$ in the text, is directly obtained from that of $\hat{B}$ by evolution with the free massless field dynamics).  

Another possibility is to treat the dynamics directly in a semiclassical or effective approximation. This alternative is better reached in our formalism by simply identifying $\hat{\cal H}_{2I}$ with zero, so that its evolution operator is the identity, and all of the interaction operator is maintained in the remaining part, that we call $\hat{\cal H}_{3I}$.

An intermediate possibility has been discussed in detail in the text. This possibility applies in a sector of the homogeneous FLRW model with a scalar field in which the integrated contribution of $8\pi G\gamma^2 W V^2/(3 \Omega_0)$ [and optionally also the ratio $8\pi G\gamma^2 W V^2/(3 \Omega_0^2)$] is small compared to the unit. In that sector, one can approximate the evolution generated by $\hat{\cal H}_{2I}$, truncating it at certain order in the potential. We have implemented this idea and computed explicitly the truncation of the evolution at second order in the potential. Obviously, one can combine this truncation in certain intervals of the evolution with other approaches, like the commented semiclassical or effective treatment, when the system moves dynamically to another region where the previous approximation is not reasonable anymore.

The last step in our analysis consists in approximating the remaining part of the interaction evolution by a semiclassical or effective counterpart. This approximation is expected to be valid as far as the FLRW state remains peaked, under the remaining interaction dynamics, on the corresponding semiclassical or effective trajectory. 

Our discussion has been applied in detail to the modifications of the MS equations deduced with a hybrid formalism in Ref. \cite{hybr-inf4}, and to a quantization of the FLRW geometry along the lines of LQC; nonetheless, we want to emphasize that the treatment is easily adaptable to other quantization approaches in homogeneous cosmology, as well as to modified MS equations obtained with other formalisms or prescriptions, as it is the case of different factor orderings, like the alternate one discussed in Ref. \cite{hybr-inf3} in order to establish an even more direct connection between the hybrid and the dress metric formulations. In those other formalisms and prescriptions, our methods can be applied in a similar way to deal with the evolution of the FLRW states, and hence be able to compute the expectation values that appear in the corresponding modified MS equations. The operators that provide the expansion of the generator of this evolution in powers of the field potential (namely, our operators ${\hat{\mathcal{H}}}_0^{(F)}$, $\hat{B}$, and $\hat{C}$) adopt different forms and representations in those other approaches, but we can treat their dynamics along the lines explained in this work. In particular, the free massless field dynamics is also analytically integrable in some other quantizations, for instance in geometrodynamics, and when this integration cannot be performed analytically, one can proceed to cope with it numerically, as we have already commented.

It is worth insisting that, in previous analyses in which the modifications to the MS equations have been treated effectively or semiclassicaly, like e.g. in Ref. \cite{javidani} for the hybrid approach to LQC, in Ref. \cite{ivans} for the dress metric approach, or in Ref. \cite{kieferdavidmanuel} for geometrodynamics, one can go beyond those approximations, keeping further quantum corrections in the computations, with the procedure that we have put forward here. Moreover, even for quantization approaches where one does not get directly a representation of the various operators that have appeared in our discussion, but rather a prescription for the calculation of expectation values, as it would be the case for path-integral formulations of quantum cosmology, one can view our analysis as a guideline, showing how to reformulate those expectation values as series of interaction terms and, possibly, how to truncate them.

An additional comment refers to our assumption of unitarity in the evolution of the FLRW states with respect to their $\phi$-dependence. In fact, revisiting the calculations of Ref. \cite{hybr-inf4}, one can check that the derivation of the MS equations \eqref{MSeqhybrid} may be extended to the situation when this assumption does not hold [though the first term in Eq. \eqref{constperturb} gets multiplied by the norm of the state, which is not preserved anymore in the evolution]. However, the unitarity of the free massless field dynamics is important for the introduction of the interaction picture, and the same occurs with the unitarity of the evolution operator $\hat{U}_{2I}$ if one wants to adopt the additional picture discussed in Sec. \ref{sec:potential}. In this respect, the unitarity of the remnant part of the dynamics is not necessary. Moreover, one can generalize the definition of the interaction picture as long as the evolution operator that is involved in the change is invertible (so that one gets a one-to-one correspondence between Hilbert spaces). In order to do this, one must substitute the adjoint of the dynamical operator by its inverse in formulas like Eqs. \eqref{intstate}, \eqref{operatorint}, or \eqref{2intpicture}. One can see that, as a result of this substitution, the relation between expectation values in different pictures [like Eqs. \eqref{expvalueint} and \eqref{expvalues2int}] varies slightly, because the operator in the original representation gets multiplied on the left by an additional factor: the composition of the operator that changes the picture with its adjoint. For instance, Eq. \eqref{expvalues2int} becomes $\langle \hat{A}_I\rangle_{ \chi_I}= \langle {\hat{U}}_J^{\dagger} {\hat{U}}_{2I}^{\dagger} {\hat{U}}_{2I}\hat{A}_J {\hat{U}}_{J} \rangle_{\chi_0} $. Finally, if the operator that should determine the change of picture were not invertible, one might still try and restrict all considerations to a subspace where it had this property.

A further comment concerns the relation between the evolution time provided by the homogeneous scalar field and the conformal time that appears in our MS equations. This relation is given by Eq. \eqref{phiconftime}. If the calculation of the expectation value of the generator of the homogeneous evolution $\hat{\cal H}_0$ becomes too complicated in practice when the field potential is present, we can always adopt the splitting of this generator explained above in terms of the generator for the free massless case, the contribution with lowest powers of the potential, and a remnant. The computation of the expectation value of the first two parts is doable. Then, one can just approximate the expectation value of the remnant by means of a semiclassical or effective approximation (or even ignore it in certain cases).

We plan to implement the above procedure in order to study the consequences of quantum corrections to the MS equations in the power spectrum of the CMB, along similar lines as in Refs. \cite{javidani,ivans,kieferdavidmanuel}, suitably modified. In the region of FLRW states and initial conditions for the perturbations where no important quantum effects are expected, we should recover the classical predictions. This will serve as a first test for the treatment proposed here. But, beyond that region, a numerical analysis starting with the steps that we have described would allow to check the validity of the semiclassical and/or effective approximation adopted in previous studies, and reveal quantum phenomena that might have been ignored in those previous works. The ultimate goal is to develop tools that allow us to predict not only the qualitative kind of corrections to the CMB spectrum that may arise from quantum geometry effects, but also to improve our control on the quantitative predictions.

\section*{Acknowledgments}
This work has been supported by MICINN/MINECO Project No. FIS2014-54800-C2-2-P from Spain. M. M-B acknowledges financial support from the Netherlands Organisation for Scientific Research (NWO) (Project No. 62001772). The authors are thankful to D. Mart\'{\i}n-de Blas and J. Olmedo for comments.

\appendix

\section{Quadratic contributions of the potential}

In this appendix, we calculate the path-ordered integral of the square of $W\hat{B}_I$ and the integral of $W^2\hat{C}_I$. These integrals appear in the quadratic contributions of the potential to the expression \eqref{quadraticAJ} of the operators in the J-interaction picture. We employ formulas \eqref{sLQCtheta} and \eqref{Boperator}, together with Eq. \eqref{Omega0hat}, for the definition of the operators $\hat{B}$ and $\hat{C}$, and we use the transformation \eqref{freeevol} to implement the change to the interaction picture. 

Let us first give the expression of the single integral. We introduce the notation
\begin{eqnarray}
\hat{\theta}_c^{(m)}(\phi)&=& \frac{1}{2^{m+2}} \cosh{\bigg( 8\sqrt{3\pi G} \hat{x}(\phi)   \bigg)}\nonumber\\
&+&\cosh{\bigg( 4\sqrt{3\pi G} \hat{x}(\phi)   \bigg)}, \label{cthetaoperators} \\
\hat{\theta}_s^{(m)}(\phi)&=& \frac{1}{2^{m+2}} \sinh{\bigg( 8\sqrt{3\pi G} \hat{x}(\phi)   \bigg)}\nonumber \\
&+& \sinh{\bigg( 4\sqrt{3\pi G} \hat{x} (\phi)   \bigg)},
\label{sthetaoperators}
\end{eqnarray} 
for any non-negative integer $m$, as well as the operators
\begin{align}
	\hat{G_1}(\phi)&= \frac{3\phi^5}{20} + \frac{\phi^4 \hat{\theta}_s^{(1)}(\phi)}{4\sqrt{3\pi G}}    {\rm sign}(\hat{\pi}_x) -  \frac{\phi^3\hat{\theta}_c^{(2)}(\phi)}{12 \pi G}   \nonumber \\
	& + \frac{3\phi^2\hat{\theta}_s^{(3)}(\phi) }{16 (3\pi G)^{3/2}}  {\rm sign}(\hat{\pi}_x) - \frac{\phi \,\hat{\theta}_c^{(4)}(\phi)}{96\pi^2 G^2}   \nonumber \\
	&+ \frac{3\,\hat{\theta}_s^{(5)}(\phi)}{128(3\pi G)^{5/2}}   {\rm sign}(\hat{\pi}_x),
	\label{Goperator1} \end{align}  
\begin{align}
	\hat{G_2}(\phi)&= \phi^4\hat{\theta}_c^{(0)}(\phi)   - \frac{\phi^3 \hat{\theta}_s^{(1)}(\phi)}{\sqrt{3\pi G}}   {\rm sign}(\hat{\pi}_x) +\frac{\phi^2 \hat{\theta}_c^{(2)}(\phi) }{4 \pi G}   \nonumber \\ &-   \frac{3\phi\, \hat{\theta}_s^{(3)}(\phi)}{8(3\pi G)^{3/2}}   {\rm sign}(\hat{\pi}_x) + \frac{\hat{\theta}_c^{(4)}(\phi)}{96 \pi^2 G^2}  .
	\label{Goperator2} \end{align}
A tedious calculation leads then to the result
\begin{eqnarray}
	\int^{\phi}_{\phi_0}&& d{\tilde \phi} W^2(\tilde \phi) \hat{C}_I(\tilde \phi)= \left(\frac{\pi G  m^2\Delta \gamma^2}{3}\right)^{2}  \nonumber \\
	\times&& \bigg(    |\hat{\pi}_x|^{1/2} \left\{\hat{G}_1(\phi)-\hat{G}_1(\phi_0)\right\}|\hat{\pi}_x|^{1/2}  \nonumber \\
	+&&  \frac{ i}{2} |\hat{\pi}_x|^{-1/2} \left\{\hat{G}_2(\phi)-\hat{G}_2(\phi_0)\right\}|\hat{\pi}_x|^{1/2} \bigg) .
	\label{intW2CI}
\end{eqnarray} 

Let us now calculate the double integral. We get 
\begin{eqnarray}
\int^{\phi}_{\phi_0} && d{\tilde \phi} W(\tilde \phi)\hat{B}_I(\tilde \phi) \int^{{\tilde \phi}}_{\phi_0}  d{\bar \phi} W(\bar \phi) \hat{B}_I(\bar \phi) = 2 \Bigg(\frac{\pi G m^2\Delta \gamma^2}{3}\Bigg)^2 \nonumber \\
\times&& \Big( |\hat{\pi}_x|   \Big\{ \hat{F}(\phi)-\hat{F}(\phi_0)\Big\}^2|\hat{\pi}_x| 
+ 4i \sqrt{3\pi G}    \hat{K}(\phi)       \hat{\pi}_x \Big),
\label{W2BI2int}
\end{eqnarray}
where $\hat{F}(\phi)$ is the operator defined in Eq. \eqref{Foperator}, and we have called
\begin{eqnarray}
 &\hat{K}&(\phi)= \int^{\phi}_{\phi_0} d{\tilde \phi} \sinh{\bigg(4\sqrt{3\pi G}\hat{x}({\tilde \phi})\bigg)}  {\tilde \phi}^2 \nonumber\\
 &\times&  \int^{{\tilde \phi}}_{\phi_0} d{\bar \phi}  \cosh^2{\bigg(2\sqrt{3\pi G}\hat{x}({\bar \phi})\bigg)}{\bar \phi}^2.
\label{Koperator}
\end{eqnarray}
This integral can be done exactly. Although long, we include the result. We get $\hat{K}(\phi)= \hat{\kappa}(\phi)-\hat{\kappa}(\phi_0)$, with
\begin{eqnarray}
&10& (48 \pi G)^3 \hat{\kappa}(\phi)= \nonumber\\
&-&  64 \sqrt{3 \pi^3 G^3} \phi^3 (72\pi G \phi^2+5) {\rm sign}(\hat{\pi}_x)  \nonumber\\
&+& 5 \bigg(4 (12\pi G \phi^2 +1)^2-3\bigg)\sinh{\bigg(8\sqrt{3\pi G} \hat{x}(\phi)\bigg)}   \nonumber\\
&-&  40\sqrt{3\pi G} (24\pi G  \phi^3+\phi ) \cosh{\bigg(8\sqrt{3\pi G}\hat{x}(\phi)\bigg)} {\rm sign}(\hat{\pi}_x) \nonumber\\
&-& 40 \sinh{\bigg(4\sqrt{3\pi G}\hat{x}(\phi)\bigg)} \bigg\{ 
24\pi G \phi \phi_0 \cosh{(4\sqrt{3\pi G}\hat{x})} \nonumber\\ 
&-&  2\sqrt{3\pi G}  \phi ( 24\pi G  \phi_0^2 +1 ) \sinh{(4\sqrt{3\pi G}\hat{x})} {\rm sign}(\hat{\pi}_x) \nonumber\\ 
&+& 24\pi G \phi(20 \pi G\phi^3-8 \pi G\phi_0^3+5 \phi) + 5 \bigg\}\nonumber\\
&+& 20  \cosh{\bigg(4\sqrt{3\pi G}\hat{x}(\phi)\bigg)} \bigg\{ 8\sqrt{3\pi G}  (96 \pi^2 G^2 \phi^5 \nonumber\\ 
&-&  96 \pi^2 G^2 \phi^2 \phi_0^3+40 \pi G \phi^3- 4\pi G \phi_0^3+ 5\phi ){\rm sign}(\hat{\pi}_x) \nonumber \\
&+&4\sqrt{3\pi G} \phi_0 (24\pi G \phi^2+1) \cosh{(4\sqrt{3\pi G}\hat{x})}{\rm sign}(\hat{\pi}_x) \nonumber\\
&-&(24\pi G \phi^2 + 1) (24\pi G \phi_0^2+1) \sinh{(4\sqrt{3\pi G}\hat{x})}  \bigg\}  .\end{eqnarray}

\end{document}